# AID: Accuracy Improvement of Analog Discharge-Based in-SRAM Multiplication Accelerator

Saeed Seyedfaraji, Baset Mesgari, Semeen Rehman
Vienna University of Technology (TU-Wien), Vienna, Austria
{saeed.seyedfaraji, baset.mesgari, semeen.rehman}@tuwien.ac.at

**Abstract**— This paper presents a novel circuit (AID) to improve the accuracy of an energy-efficient in-memory multiplier using a standard 6T-SRAM. The state-of-the-art discharge-based in-SRAM multiplication accelerators suffer from a non-linear behavior in their bit-line (BL, BLB) due to the quadratic nature of the access transistor that leads to a poor signal-to-noise ratio (SNR). In order to achieve linearity in the BLB voltage, we propose a novel root function technique on the access transistor's gate that results in accuracy improvement of on average 10.77 dB SNR compared to state-of-the-art discharge-based topologies. Our analytical methods and a circuit simulation in a 65 nm CMOS technology verify that the proposed technique consumes 0.523 pJ per computation (multiplication, accumulation, and preset) from a power supply of 1V, which is 51.18% lower compared to other state-of-the-art techniques. We have performed an extensive Monte Carlo based simulation for a 4x4 multiplication operation, and our novel technique presents less than 0.086 standard deviations for the worst-case incorrect output scenario.

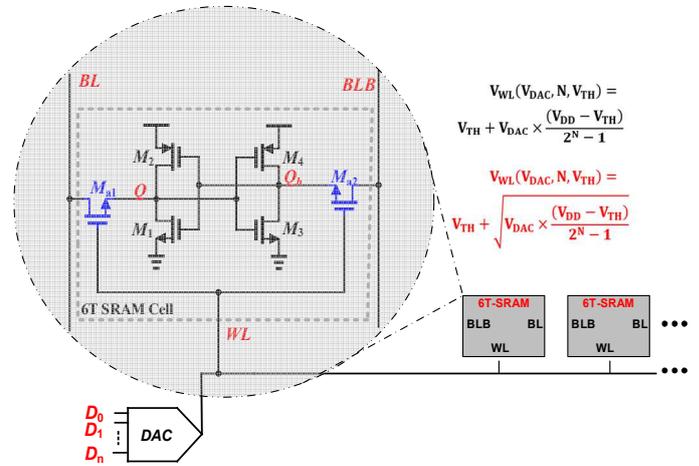

Fig. 1: Standard single bit 6T-SRAM memory cell. Appling different word-line voltage, semi-linear (conventional techniques), and root function (our proposed technique)

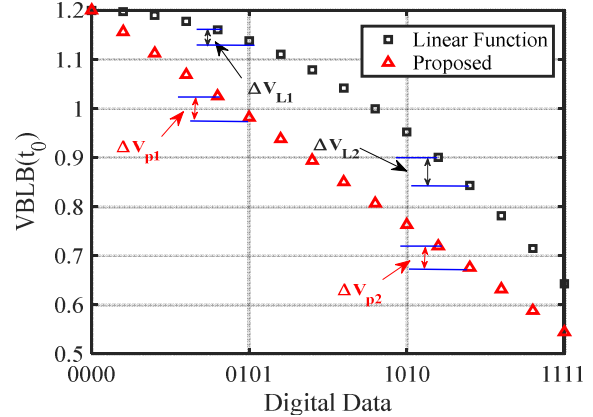

Fig. 2: Non-linear behavior of BLB voltage and the effect of our proposed linear root function technique.

## I. INTRODUCTION

Machine Learning (ML)/Artificial Intelligence (AI) based algorithms are getting increasingly in demand and are widely deployed in today's smart applications (e.g., automotive, hospitals and wearable healthcare, etc.) that enable high quality of life, safety, and healthcare [1]. The underlying basic building block of AI/ML based algorithms are the multi-bit multiplication and addition operation, fulfilling the demands for faster and accurate results. However, these operations incurs high energy-/power costs [1]. A large part of the total system's energy-/power consumption is associated with the data movement between memory and processing unit [2]–[4]. It is reported in [1] that, several millijoules of energy per operation is required in a conventional Von-Neumann processing model. In order to address this problem, Process-In-Memory (PIM) concept has emerged as a promising way to reduce the energy consumption associated with the data movement (between memory and processing unit), by equipping memories with substantial processing capabilities while exploiting their analog and physical properties to accelerate the computation.

Before deploying the PIM technique, it is imperative to first evaluate the characteristics of various memory technologies w.r.t power, energy and latency for the targeted system. Among all the memory technologies, SRAM attracts our attention due to its high sub-bank divisibility feature [3], that enabled us to apply our PIM technique in a fine-grained manner using sub-banks [3], [5]–[7] inside the SRAM. Such approach enables high parallelism during computation, which is primarily due to the wideband internal connection and abundance of sub-banks that ultimately increases the performance by several orders of magnitude. SRAM-based PIM techniques are mainly divided in two categories: a) boolean vector (digital bitwise) calculation , and b) analog discharge behavior interpretation. The first category is based on activating two word-lines (WL) concurrently, and using an adjusted sense amplifier (SA) to identify the output. Combining such complicated SA's with extra circuitries as peripheral devices can enable complex computation such as addition [7]–[14]. These approaches increase latency, area, and energy consumption of the final circuit. In the second category, the result of the mathematical operation is modulated (encoded) based on the discharge behavior (voltage drop) of the Bit-line-bar (BLB) of the SRAM cell. In order to enable multiplication, the produced voltage of each BLB should be weighted and combined with other BLB's, proportional to the digital input weight of the operands (i.e., the LSB with the weight of "1" and MSB with weight of $2^{(n-1)}$) [15]. To the best of our knowledge, the state-of-the-art approaches uses the standard 6T-SRAM cell as the primary memory unit, because it occupies minimum area and consumes less power. One of the commonly used approaches to modulate the digital input to the analog counterpart is via using the amplitude and pulse-width (PW) of the access transistor (as is shown in Fig. 1. The modulated data is applied to the SRAM cell via the gate port of the access transistor, either using different amplitude and PW, or a combination of both techniques [10]–[12], [14], [15]. Ranging the amplitude of the gate port of the access transistor might introduce some non-linearities in the discharge behavior of BLB's, leading to an increase in the Bit Error Rate (BER) of the sampled data. To overcome this problem, the authors in [15] apply the digital input data to the access transistor using a semi-linear function of the digital codes (depicted by the first equation in Fig. 1). This function led to a non-uniformly spaced analog voltage for the BLB discharge behavior, as illustrated in Fig. 2. Consequently, due to the non-linear nature of the access transistor, the voltage will drop having non-equal differences, that will result in high BER. Fig. 2 shows the digital number versus the interpolated analog counterpart at the BLB port, based on the linear function at the gate of the access transistor. For example, for a 4 bits number, interpolated analog values from "0000" to "0101" are not sufficiently spaced, which means Analog to Digital Converter (ADC) cannot distinguish between these six numbers. In Fig. 2, the $\Delta V_{L1}$ is smaller than the $\Delta V_{L2}$; as a result, the probability of converting the expected multiplication value to the true digital number without an error is significantly low, as compared to the numbers assigned to the $\Delta V_{L2}$. Although in this work [15], a binary-weighted charge sharing technique is also utilized to modify the accuracy of the computation, but the required bit margin is still not sufficiently spaced, and a non-zero static current due to the usage of the pre-charge circuit is imposed. This static current might not be ignored when the size of the array is relatively large.

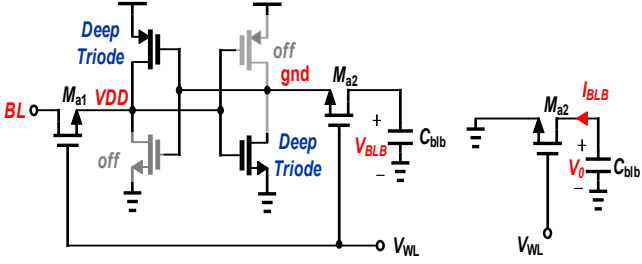

Fig. 3: Equivalent circuit diagram of discharge behavior of BLB.

**Our Novel Contributions in a nutshell are:** 1) we have performed an extensive analysis to identify the non-linear behavior of voltage drop in the state-of-the-art methods [15][16] which leads to BER. 2) We have proposed AID circuit, which force the BLB voltage to drop following a uniform step to enhance computation accuracy. By doing so, the dynamic range of the BLB voltage is increased in such a way that the $\Delta V_{p1}$ becomes equal to the $\Delta V_{p2}$ (see Fig. 2). 3) we have proposed a root function (represented in the equation 2, Fig. 1) circuit that will be applied to the gate of the access transistor in our AID circuit. By utilizing this technique as word-line data, the compression of interpreted voltage is removed, and the digital input bits are modulated in a linear manner, which can drastically enhance the BER. 4) Finally, we have performed an extensive evaluation of our proposed method to properly confirm the improvement of our circuit. To the best of authors' knowledge, AID is the first circuit that eliminates the non-linearity in the input of the ADC deployed inside the in-memory multipliers, and significantly reduces it's output error rate. The rest of the paper is organized as follows: Section II.A provides an overview of the SRAM operation to understand how various circuit level parameters are selected in order to perform an analog multiplication, and the accuracy of the proposed structure is also formulated. Our novel multi-bit multiplication design is provided in section III, followed by section VI presenting the results and experimental setup, where the AID circuit level simulation, the process variation analysis and the state-of-the-art comparisons are provided. We conclude this manuscript in section V.

## II. ANALYTICAL ASPECTS OF ANALOG MULTIPLICATION IN THE STANDARD 6T-SRAM CELL

### A. SRAM Read, Write, and Multiplication Operation

Fig. 3 illustrates the standard 6T-SRAM, a fundamental building block of a static memory cell. In order to perform the conventional read and write operations in a 6T-SRAM cell, at first the preparations for setting up the circuit has to be accomplished. *For a read operation*, both bit-lines i.e., *BL* and *BLB* are initially floating high, which means that the gate voltage of the access transistors (i.e., $M_{a1}$ and $M_{a2}$ in Fig. 3) are already discharged to the ground (*WL* is almost zero), assuming that $Q$ is equal to zero, and therefore, $Q_b$ is subsequently *VDD*. *BL* and *BLB* should be recharged to *VDD*, and therefore, the *WL* is enabled by applying the *VDD* voltage to the gates of $M_{a1}$ and $M_{a2}$. In such a case, the gate-source voltage of $M_{a2}$ is almost zero (*WL=VDD* and $Q_b$=*VDD*). As a result, there is no significant current passing, and *BLB* and $Q_b$, are remained unchanged. Unlike *BLB*, the *BL* is pulled down through $M_{a1}$ and $M_1$ (see Fig. 1), leading to a reading operation of the stored value ($Q = 0$) from the cell. Likewise, when *VDD* is stored in the cell, *BL* should remain unchanged, but *BLB* must be discharged to zero. As discussed, for the reading operation of both zero and *VDD* from the cell, *BL* and *BLB* were pre-charged to *VDD* at the beginning. However, *for a write operation*, different directions of the charging of *BL* and *BLB* are required. To clarify, for writing *VDD* into the cell, *BL* should be pre-charged to *VDD*, while *BLB* is pulled down to zero and *WL* is then increased.

In order to perform analog multiplication, in this paper we intend to use the read operation of conventional SRAM cell when $Q = VDD$, and $Q_b =$ "0". We have accomplished the dot product by modulating the input digital data $Din$ ($D_{n-1}, D_n …, D_1, D_0$) into amplitude of the applied signal to the *WL*. We will discuss the in-SRAM analog multiplication in detailed next.

### B. Quantitative Describtion of Mixed-Mode Multiplication Using a Standard SRAM cell

In the read operation, for performing multiplication, instead of reading the specific digital data, a resultant product between two multi-bit words can be modulated in the discharge behavior of *BLB*. In order to provide an in-depth insight into discharge behavior of *BLB*, we have performed an extensive analysis to understand the impact of the *WL*'s amplitude on the output BER. At the first phase of multiplication, *WL* will be discharged to the ground (disable mode), and we write *VDD* into SRAM cell, which means $Q = VDD$ and $Q_b = 0$. At the second step, *BL* and *BLB* should be pre-charged to *VDD* for a read operation. Fig. 3 shows the preparation phase for analog multiplication. Considering Fig. 3, when $Q = VDD$ and thus $Q_b = 0$, $M_3$, $M_2$ are in the deep triode region (i.e. their channel resistance is relatively low), and $M_4$, $M_1$ are at the cut-off region.

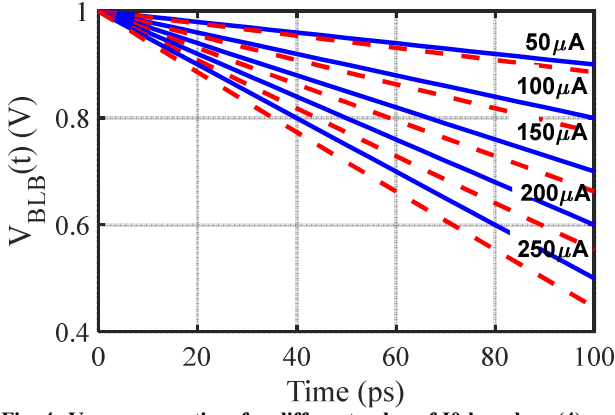

Fig. 4: $V_{BLB(t)}$ versus time for different value of I0 based on (4) and (5).

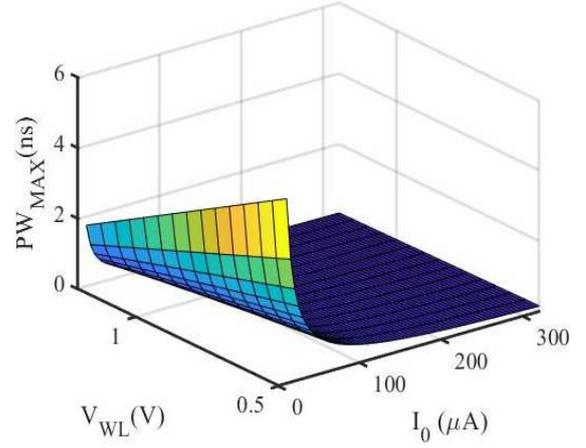

Fig. 5: Maximum sampling time of $V_{BLB}$ when $M_{a2}$ stays in the active region versus $I_0$ and $V_{WL}$ using equation (6).

Once WL is enabled, the only possibility of establishing a significant current flow is through the $M_{a2}$ and $M_3$ (see Fig. 1). Consequently, $M_{a2}$ creates a path for BLB to discharge the initial voltage of $C_{blb}$ from VDD towards the ground (0V). Based on our above discussion, the rightmost transistor (in Fig. 3) is an equivalent circuit to quantify the current flow of $I_{BLB}$. By applying Kirchhoff's Current Law, at the drain node of this transistor, the difference equation of discharge behavior of BLB can be obtained using equation (1).

$$I_{BLB} + C_{blb}\frac{dV_{BLB}(t)}{dt} = 0 \qquad (1)$$

For the current equation of the MOSFET, we first neglect the channel-length modulation, and then the drain-source current of $M_{a2}$, which is named as $I_0$, is expressed using equation (2).

$$I_0 = \frac{1}{2}\mu_n C_{ox}\left(\frac{W}{L}\right)(V_{GSa2} - V_{TH})^2 \qquad (2)$$

$$-\int_{t=0}^{t}\frac{I_0}{C_{blb}}d\tau = \int_{V_0=V_{DD}}^{V_{BLB}(t)} dV_{BLB}(\tau) \qquad (3)$$

For this equation, we assume that the transistor remains in the saturation region during the discharging time of BLB. In equation (2), $\mu_n$ is the mobility of the electron, and $C_{ox}$ is the gate-oxide capacitance of the transistor. The value W, and L expresses the channel width and length of the MOSFET transistor, respectively. $V_{TH}$ shows the threshold voltage, which is the minimum gate-to-source voltage to create a nonzero current between the sources and drain terminals. In order to consider the channel-length modulation parameter $\lambda$ in the current equation of $M_{a2}$, equation (2) needs to be multiplied by a factor (1+$\lambda V_{BLB}$). By combining equations (1) and (2), and also applying integral functions for both sides, we obtained equation (3). By exploiting the difference in equation (1) and multiplying $I_0$ by the factor of (1+$\lambda V_{BLB}$), and also solving this equation and finding $V_{BLB}(t)$, the effect of channel-length modulation can be considered, which is expressed in equation (5) that has a non-linear discharging behavior. Equation (4) and (5) show that applying an analog voltage with the different amplitude and PW to the gate of $M_{a2}$ can lead to generating a variety of currents through the transistor, and then as a result, different voltage drops of BLB terminal will be appeared. Fig. 4 illustrates the discharge behavior of BLB using 4 blue solid lines, and 5 red dashed lines. For this simulation, we assumed that VDD = 1V, $C_{blb}$ = 50 fF, $\lambda$=0.15V$^{-1}$. As an observation, a multi-bit input can be mapped as discharge voltage on the BLB within a specified time (PW), and a different value for $I_0$ (Amplitude).

$$V_{BLB}(t) = V_{DD} - \frac{\overbrace{\mu_n C_{ox}\left(\frac{W}{L}\right)}^{\beta}(V_{WL} - V_{TH})^2 t}{2C_{blb}} \qquad (4)$$

$$-\frac{I_0 t}{C_{blb}} = \frac{1}{\lambda}ln\left(\frac{V_{BLB}(t) + \frac{1}{\lambda}}{V_{DD} + \frac{1}{\lambda}}\right) \to V_{BLB}(t)$$
$$= \left(V_{DD} + \frac{1}{\lambda}\right)e^{-\left(\frac{\lambda I_0}{C_{blb}}\right)t} - \frac{1}{\lambda} \qquad (5)$$

We should emphasize that changing the $V_{WL}$ using a linear function (as presented in previous state-of-the-art approaches) cannot lead to a linear relation between the digital input data and resultant BLB voltage, which might induce a considerable error in the multiplication result. As mentioned earlier, equation (2) is valid when $M_{a2}$ is still at the saturation region. Hence, during the discharging time of BLB, we should consider a condition, so $M_{a2}$ does not face the triode region; otherwise, the BLB discharging rate gets relatively slow. Consequently, accurate sampling of the BLB voltage just before the transistor ($M_{a2}$) enters the triode region can prevent systematic errors. The accurate sampling can be done by controlling the pulse-width of $V_{WL}$. As a result, based on equation (4) together with applying saturation condition of an NMOS transistor, maximum $PW_{Max}$ is calculated using equation (6). Considering design parameters used for Fig. 4 (as mentioned earlier), a 3-D representation of equations (6) is illustrated in Fig. 5, which can lead to a precise prediction for maximum $PW_{Max}$ of voltage drop of the BLB for different currents and $V_{WL}$ voltages while the access transistor still works at the active region. This curve implies that, the more current passes via access transistor, the less sampling time will be needed.

$$V_{DMa2} \geq V_{GMa2} - V_{TH} \to$$
$$PW_{Max} \leq \frac{C_{blb}}{I_0}(V_{DD} + V_{TH} - V_{WL}) \qquad (6)$$

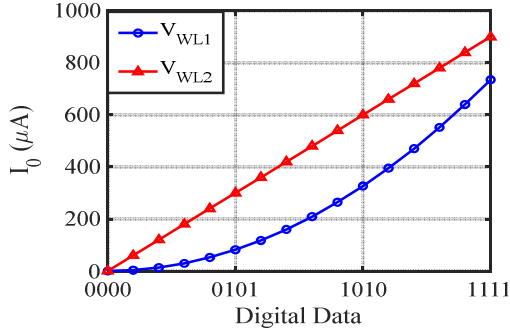

Fig. 6: $I_0$ versus digital Data using equations (7) and (8).

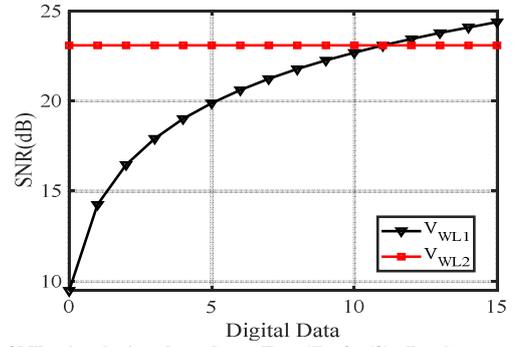

Fig. 7: SNR simulation based on Eq. (7) & (8). In the propsed method ($V_{WL2}$) 10.77 dB SNR improvement is achieved.

### C. Mapping The Digital Data To The Analog Voltage of $V_{WL}$ And Linearity Issue

For an N×N multiplication, where N is an arithmetic number ($N \epsilon W$), ignoring the repetitive instances we would need to generate a (0.5×N×(N-1) +1) linearly spaced transistor current by controlling $V_{WL}$, thereby controlling the discharge rate of BLB. For example, in [15], a linear function generates the $V_{WL}$, formulated in equation (7). Using (7) as word-line voltage ($V_{WL1}$) cannot produce a linear current flowing through $M_{a2}$, since in this case, $V_{BLB}$ is changing based on the quadratic behavior of the $V_{DAC}$. As a solution, in this paper, we propose to use a root function for $V_{WL2}$, which is expressed in (8). $V_{DAC}$ refers to the controlling voltage of $V_{WL}$, which is usually accomplished using a digital to analog converter (DAC).

$$V_{WL1} = V_{TH} + V_{DAC} \times \frac{(V_{DD} - V_{TH})}{2^N - 1} \quad (7)$$

$$V_{WL2} = V_{TH} + \sqrt{V_{DAC} \times \frac{(V_{DD} - V_{TH})}{2^N - 1}} \quad (8)$$

Considering the parameter assumptions of Fig. 4, and also equations (2), (7), and (8), $I_0$ is plotted for two different cases ($V_{WL1}$ and $V_{WL2}$) versus digital data in Fig. 6. Based on this figure, by applying $V_{WL2}$ at the gate of $M_{a2}$, $I_0$ has obtained a linear proportionality against the digital input data, while $V_{WL1}$ has led to a non-linear behavior.

### D. Theortical calculation of Signal to Noise ratio of Analog multiplication

In this section, we quantify the analog multiplier's accuracy employing SNR calculation, at the BLB port. As we discussed previously, utilizing a linear function at the gate of access transistor would introduce BER degradation. To formulate the BER degradation we have used SNR which can be easily expressed using circuit parameters. There is an inverse relation between SNR and BER, so if BER increases, the SNR decreases and vice-versa. We consider two different cases based on Fig. 2. For SNR calculation we use equation (9), $P_{noise}$ can be approximated by assuming the integrated noise variance of a parallel RC network, which is $\delta^2 = KT/C$. K is the Boltzmann's constant, T expresses the absolute temperature in Kelvin.

$$SNR = 10\log_{10}\left(\frac{P_{Signal}}{P_{noise}}\right) \quad (9)$$

$P_{signal}$ has to be calculated based on two successive BLB voltages due to the change of DAC digital input ($V_{DACi}$ and $V_{DACi+1}$). By substituting $V_{WL1}$ (7) and $V_{WL2}$ (8) into equation (4) and applying algebraic simplification, voltage differences for both cases can be achieved using equations (10) and (11), respectively. The $t_0$ is the sampling time of BLB voltage, which its upper limit can be calculated using Fig. 5, and equation (6).

$$\Delta V_{BLB V_{WL1}} = \frac{\beta t_0}{C_{blb}}\left(\frac{(V_{DD} - V_{TH})}{2^N - 1}\right)^2 ((V_{DACi})^2 - (V_{DACi+1})^2) \quad (10)$$

$$\Delta V_{BLB V_{WL2}} = \frac{\beta t_0}{C_{blb}}\left(\frac{(V_{DD} - V_{TH})}{2^N - 1}\right)(V_{DACi} - V_{DACi+1}) \quad (11)$$

By dividing $\Delta V^2_{BLB}$ to the $\delta^2$ and utilizing equation (9) the SNR of multiplication can be obtained versus the input digital data. To show the efficiency of our proposed method in applying a non-linear function in the word-line port, a simulation, based on Eq. (9-11) and assuming that $VDD = 1V$, $C_{blb} = 50$ fF, and $t_0=50$ ps has been done, as shown in Fig. 7. By applying the $V_{WL1}$ which is a linear function of the input digital data, has led to a non-uniform spaced analog voltage for BLB discharge behavior illustrated in Fig. 2. However, in the propose method, by forcing the BLB voltage to follow a linear trend via utilizing root function voltage for the gate of the access transistor, it leads to a relative accuracy improvement of up to10.77 dB average SNR.

### III. MULTI-BIT CALCULATION IN 6T-SRAM

We employ the standard 6T-SRAM cell as a single-bit multiplication unit as depicted in Fig. 1. Suppose we tend to do a multiplication of 4-bit $D_{in}$ and 4-bit stored data in SRAM cells, named as $J_s$ ($J_3…J_0$) in this paper. If the stored data needs to have an 8-bit resolution, we should use eight single-bit SRAM cells as an array, while for $D_{in}$, the required resolution is coded using the amplitude of $V_{WL}$. Fig. 8, label **(a)**, shows the architecture of the proposed in-memory 4-bit $D_{in}$ multiply by 4-bit $J_s$ multiplier. It consists of four standard 6T- bit-cell and read and write circuitry for two operating modes. In the SRAM mode, the operation is a standard read/write of digital data, but in multiplication mode, a sampled product of analog multiplication is generated. As the first step, which is discussed earlier, $J_s$ is stored in the 4-bit array during $T_{WEN}$, $BL$, and $BLB$ reaching $VDD$ for a multiplication operation, exactly right after data storing period which is called $T_{pre}$ (see Fig. 8, label **(b)**). However, if "0" is stored, there will not be an enough current flowing; consequently, $BLB$ remains unchanged and stays close to $VDD$. The 4-bit $J_s$ is stored inside four-unit cells of a 6T-SRAM in a column as highlighted in Fig. 8, label (a). MSB's are stored in the left cells, while LSB's are located in the right cells. In order to conduct an accurate analog multiplication while considering bit significance for $D_{in}$, we use two techniques namely a non-linear DAC in conjunction with a charge sharing circuit, as shown in Fig. 8 (label (a), green highlighted). Our charge sharing method does not need a static current in contrast with other state-of-the-art approaches, especially in [15], which has used the pre-charge circuit PW control, imposing a cost of static current. As it shown in Fig. 8, the PW of each path is controlled using two switches, which are working complementary; for example, when $S_3$ is connected, $S_{3b}$ is open. By doing so, the PW of BLB can be adjusted based on bit significance. As a result, current flowing time ($T_0$) has been assigned to MSB while LSB has a higher discharging time ($8 \times T_0$).

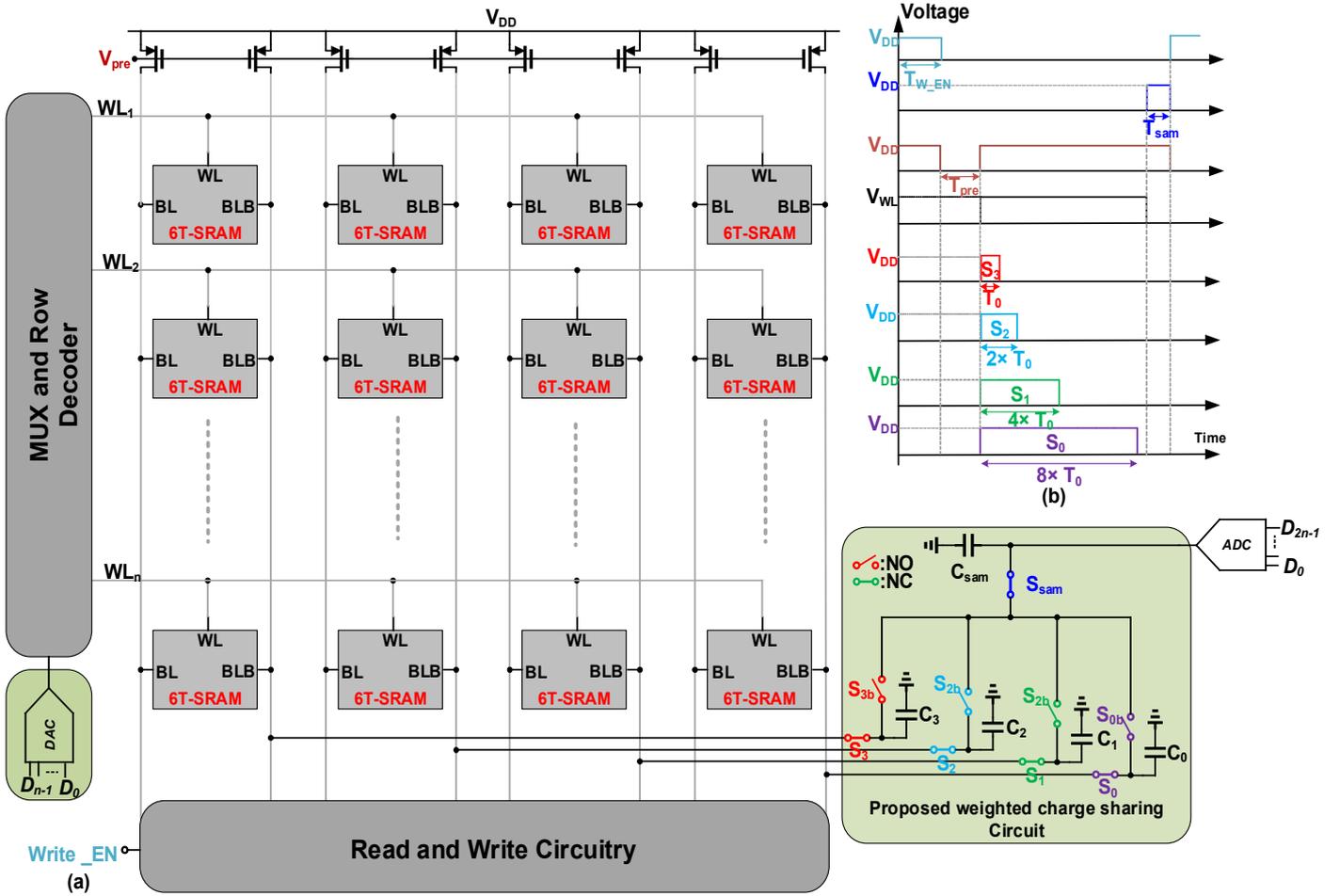

Fig. 8: In-Memory multiplication operation 4-bit Din multiply by 4-bit Js; Digital data (Din) is coded using the amplitude of word-line. we propose two techniques namely a nonlinear DAC (root function) in conjunction with a charge sharing circuit which both are highlighted in green.

It is worth mentioning that, based on equation (4), (8), using our weighted charge sharing method can lead to a proportional discharge behavior of *BLB* which is essential for providing high accuracy multiplying results. In addition to the time coding of input data provided by the proposed charge sharing circuit, to achieve a higher resolution of *BLB* discharging rate, the amplitude of $V_{WLs}$ has been selected based on the bit significance of $D_{in}$ as we described them in the previous section. Consequently, the amplitude of word-line terminals should be chosen based on the fact where $V_{WL0} > V_{WL1} > V_{WL2} > V_{WL3} > V_{TH}$. To provide stable *BLB* data for using ADC at the end of ($T_{WEN} + T_{pre} + 8 \times T_{on} + T_{sam}$) = $T_{MU}$ time frame, a sample and hold circuit is needed as exhibited in Fig. 8, label (a). When $S_{sam}$ is opened, the $V_{BLBS}$ will be sampled and delivered to the ADC.

## IV. EXPERIMENTAL SETUP OF OUR PROPOSED ANALOG MULTI-BIT MULPLICATION IN THE 65NM CMOS TECHNOLOGY

The design depicted in Fig. 8. (a), is implemented at the circuit level, in 65 nm CMOS technology with a supply voltage of 1 V. The circuit parameters such as the size of the transistors, word-line voltages as well as PW of charge sharing circuit are optimized using SPECTRE transient simulation and according to equations (1) - (11). This optimization is aimed to minimize $T_{MU}$, power consumption, and occupied area, such that the accuracy of multiplication is enhanced. Fig. 9 shows the discharge behavior of BLB versus time for a different value of $V_{WL}$. As a comparison with Fig. 4, the simulated result can prove that the efficiency of our mathematical analysis is highly precise, more specifically in the linear region. As a result, sampling $V_{BLB}$ at a specific time is referred to input digital value. Assuming a 4-bit $D_{in}$, $V_{WL}$=0.6V can be interpreted as a binary value of "1111" while 1V is "0000". Random variations of employed transistors such as threshold voltage, gate oxide thickness, and various mobility are the significant problems of circuit design performance degradation of analog multiplication. Therefore, in this paper, a 1000 points Mont-Carlo simulation is done, while considering process and mismatch to characterize the accuracy of 4-bit $D_{in}$ multiply by 4-bit $J_s$. As mentioned, 4-bit $D_{in}$ is coded into amplitude of the word-line while $J_s$ are directly stored inside each 6T-SRAM. The worst-case standard deviation is less than 0.086 for case of (15×15), which is illustrated in Fig. 10. To evaluate performance improvement of our work (AID) compared to other topologies, Table 1 presents the MAC energy and standard deviation of the our's and the state-of-the-art techniques.

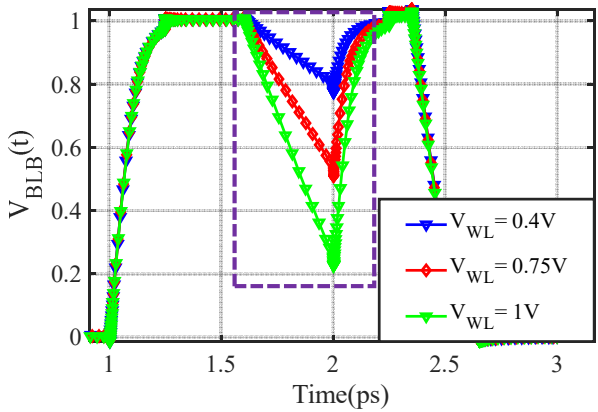

Fig. 9: *Simulated $V_{BLB}(t)$* vs. time for different $V_{WL}$. Compared to Fig. 4 simulation follows the theoritical equations.

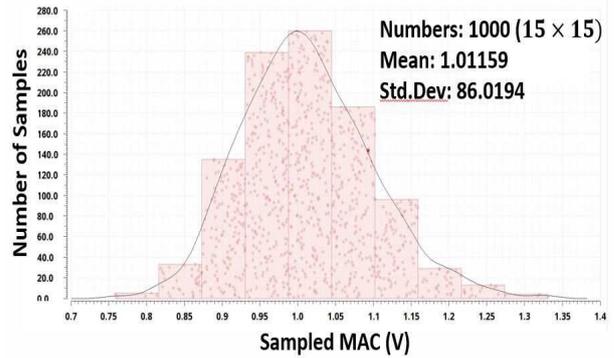

Fig. 10: The worst-case standard deviation of our proposed analog multiplication; achieved by a 1000 points Monte-Carlo simulation.

TABLE 1. COMPREHENSIVE COMPARISON OF AID AND STATE-OF-THE-ART TECHNIQUES

|  | AID | [15] | [16] | [12] | [17] | [10] |
|---|---|---|---|---|---|---|
| Tech. (nm), Voltage (V) | 65-1 | 65, 1.2 | 65, 1 | 180, 1.8 | 65, 0.925 | 65, 1.2 |
| Output bit | 4 | 4 | 8 | 5 | 4 | 5 |
| MAC energy (pJ) | 0.523 | 0.9 | 1.3 | 1.167 | 0.32 | 3.5 |
| Accuracy (STD.V) | 0.086 | 0.6 | / | / | / | / |
| Freq. (MHz) | 200 | 100 | 60-125 | / | / | 2.5 |

I. CONCLUTION

This paper presents a novel technique (AID) to improve the accuracy of an energy-efficient in-memory multiplier using a standard 6T-SRAM. Our solution is to force the BLB voltage to follow a linear, leading to a relative accuracy improvement up to 10.77 dB average SNR compared with other discharge-based topologies. Our comprehensive evaluation in a 65 nm CMOS technology verifies that the proposed technique only consumes 0.523 pJ per computation (multiplication, accumulation, and preset) from a power supply of 1V, and for the worst case, our result standard division is less than 0.086.